\documentclass[
  aps,
  pre,
  reprint,
  amsmath,
  amssymb
]{revtex4-1}
\setcitestyle{square}
\usepackage[pdftex]{graphicx}
\usepackage[pdftex,bookmarks,colorlinks,citecolor=blue,linkcolor=blue]{hyperref}
\usepackage[pdftex,svgnames,usenames,dvipsnames]{xcolor}
\usepackage[utf8]{inputenc}
\usepackage[T1]{fontenc}
\usepackage{siunitx}
\usepackage{bm}

\newcommand{\figref}[1]{Fig.~\ref{#1}}

\newcommand\ReNum{\mbox{Re}}  

\begin{document}

\title{Non-monotonic flow curves of shear thickening suspensions}

\author{Romain Mari}
\affiliation{Benjamin Levich Institute, %
City College of New York, New York, NY 10031, USA}
\author{Ryohei Seto}
\affiliation{Benjamin Levich Institute, %
City College of New York, New York, NY 10031, USA}
\author{Jeffrey F. Morris}
\affiliation{Benjamin Levich Institute, %
City College of New York, New York, NY 10031, USA}
\affiliation{Department of Chemical Engineering, %
City College of New York, New York, NY 10031, USA}
\author{Morton M. Denn}
\affiliation{Benjamin Levich Institute, %
City College of New York, New York, NY 10031, USA}
\affiliation{Department of Chemical Engineering, %
City College of New York, New York, NY 10031, USA}

\date{\today}

\begin{abstract}
The discontinuous shear thickening (DST) of dense suspensions is a
remarkable phenomenon in which the viscosity can increase by several
orders of magnitude at a critical shear rate.
It has the appearance of a first order phase transition
between two hypothetical ``states'' that we have recently identified
as Stokes flows with lubricated or frictional contacts, respectively.
Here we extend the analogy further by means of novel stress-controlled
simulations and show the existence of a non-monotonic steady-state
flow curve analogous to a non-monotonic equation of state.
While we associate DST with an S-shaped flow curve, at volume
fractions above the shear jamming transition the frictional state
loses flowability and the flow curve reduces to an arch,
permitting the system to flow only at small stresses.
Whereas a thermodynamic transition leads to phase separation
in the coexistence region,
we observe a uniform shear flow all along the thickening transition.
A stability analysis suggests that uniform shear may be mechanically
stable for the small Reynolds numbers and system sizes in a rheometer.
\end{abstract}

\maketitle

\section{Introduction}
\begin{figure}[t]
  \centering
  \includegraphics[width=0.45\textwidth]{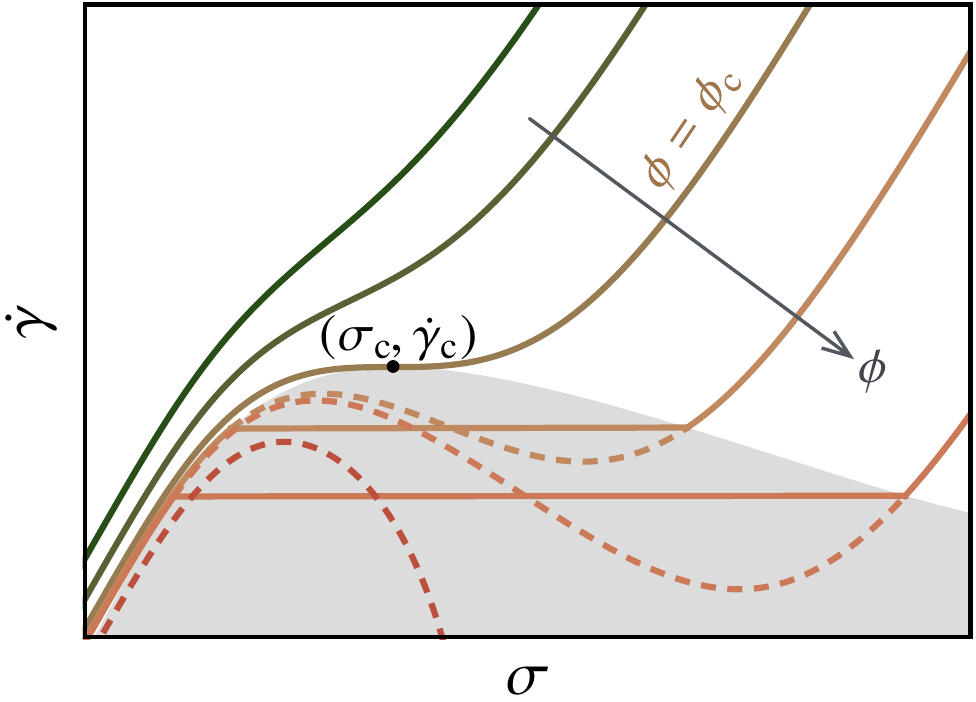}
  \caption{
    (Color online).  Sketch of the relation between the shear rate
    $\dot\gamma$ and the shear stress $\sigma$ in a shear thickening
    suspension, with increasing volume fraction $\phi$ from top to
    bottom, in solid lines.
    At low $\phi$, $\dot\gamma(\sigma)$ is a strictly monotonic
    function and the shear thickening is continuous,
    appearing as an inflection in the curve.
    As $\phi$ increases, this inflection deepens, and at $\phi=\phi_{\mathrm{c}}$
    a ``critical'' point $(\sigma_{\mathrm{c}},\dot\gamma_{\mathrm{c}})$
    appears in which $\mathrm{d}\dot\gamma/\mathrm{d}\sigma=0$.
    For $\phi>\phi_{\mathrm{c}}$ the shear thickening is discontinuous.
    The discontinuity region (with constant $\dot\gamma$) 
    is analogous to a coexistence region (light gray region). 
    (Hysteresis is sometimes observed,
      in which case the discontinuity region would be analogous to a spinodal region.)
    The discontinuity region is only accessible under stress-controlled conditions.
    The putative non-monotonic flow curves are plotted as dashed lines in the coexistence region,
    in analogy to the non-monotonic equations of state
    underlying a first order thermodynamic transition.
  }
\label{fig:analogy_transition}
\end{figure}

The shear thickening of dense suspensions is a counter-intuitive phenomenon
in which, for some range of applied shear stress,
the viscosity of a suspension increases, sometimes by orders of
magnitude~\citep{Metzner_1958,Barnes_1989,Brown_2014}.
It is observed in systems that differ widely in the shape, type or
size of the suspended solid material, or in the type of suspending
liquid; it is even argued that most dense suspensions shear thicken,
but the phenomenon may often be hidden by a yield
stress~\citep{Brown_2010,Franks_2000}.
Despite being an inherently non-equilibrium phenomenon,
shear thickening shares some strong similarities with an equilibrium phase
transition.
In \figref{fig:analogy_transition} we show schematically the relation
between the shear stress $\sigma$ and the shear rate $\dot\gamma$ as
it is commonly observed in
experiments~\citep{Egres_2005a,Fall_2008,Brown_2009,Brown_2014,Mewis_2011}.
The curves observed at different volume fractions $\phi$ are
reminiscent of the isotherms in a $P$-$V$ diagram for a system
undergoing a liquid-gas transition.
This leads to an analogy between discontinuous shear thickening (DST)
and a first order transition.


Along this line of thought, \citet{Wyart_2014} recently suggested that
a non-monotonic flow curve $\dot\gamma(\sigma)$, as represented
in~\figref{fig:analogy_transition}, underlies the discontinuity
region, just as a first order transition stems from a non-monotonic
equation of state.
(The hypothesis of an S-shaped flow curve is controversial~\citep{Pan_2014,Fall_2015}. The arguments
  of~\citet{Wyart_2014} rely on an interpolation between two hypothetical coexisting states, one frictionless and one frictional.)
For the liquid-gas transition, the S-shaped equation of state is never
observed even in a volume-controlled experiment, because it leads to a
coexistence region where the uniform system is no longer a global
minimum of the free energy and the system phase separates.
For DST, however, one cannot rely on a thermodynamic argument and we
are restricted to mechanical considerations.
For micellar systems, a mechanical stability analysis of the
``spinodal'' region (the region inside the coexistence region with
$\mathrm{d}\dot\gamma/\mathrm{d}\sigma<0$) leads to an instability
of the uniform flow towards shear banding, which is a mechanical phase
separation~\citep{Spenley_1996, Coussot_2014}.
Interestingly, even if many DST experiments show the
``thermodynamic'' behavior described in
\figref{fig:analogy_transition}, with a shear rate plateau in the
discontinuity region under stress-controlled conditions, some
experiments instead show a non-monotonic rheology~
\citep{Laun_1994, Frith_1996, Neuville_2012},
and a very recent study~\citep{Pan_2014} clearly maps out
a full S-shaped curve for a neutrally buoyant suspension of spheres, but with hysteresis and a rate dependence of the transition in the flow curve.
Some related granular flows have been studied at fixed shear
stress and showed a similar
non-monotonic $\dot\gamma(\sigma)$ relation, but only as a
transient~\citep{Grob_2014}.

Here we address these properties of DST
and non-monotonicity with numerical simulations of a model shear-thickening
suspension of hard frictional spherical particles.
The simulation model, which considers frictional contacts between
suspended particles, has recently been shown to reproduce both continuous shear thickening (CST) and
DST under rate control, as observed in
experiments~\citep{Heussinger_2013,Fernandez_2013,Seto_2013a,Mari_2014}.
Modern shear rheology experiments are often carried out under
  shear stress rather than shear rate control, but existing methods to simulate
  stress-controlled flows of dense suspensions, by introducing walls
  or feedback loops, are of limited utility and cannot be used to simulate DST.
  We describe here a novel deterministic stress-controlled scheme
  employing periodic boundary conditions that makes it possible for
  the first time to simulate the flow in the DST regime.
  We observe both the S-shaped flow curves and
  the arches suggested by analogy to a first order transition (cf. \figref{fig:analogy_transition}) \emph{in
    steady state with a uniform flow}; there is no mechanical
  instability leading to shear-banding or chaotic
  dynamics~\citep{Nakanishi_2012}, in stark contrast to a thermodynamic transition.

\section{Methods}
We simulate an assembly of inertialess frictional spheres
immersed in a Newtonian fluid under simple shear flow.
The system is binary, with radii $a$ and $1.4a$ mixed at equal volume fractions.
A stabilizing repulsive force that prevents frictional contacts between
  particles at low stresses, as used in our basic model~\citep{Seto_2013a}, is a key ingredient in shear thickening; in many shear thickening
  experiments this takes the form of an electrostatic double-layer
  force~\citep{Laun_1994}.
The equation of motion is thus simply
the force balance between hydrodynamic ($\bm{F}_{\mathrm{H}}$),
repulsive ($\bm{F}_{\mathrm{R}}$), and contact ($\bm{F}_{\mathrm{C}}$) forces
which depend on the many-body position and velocity vectors $\bm{X}$ and
$\bm{U}(\equiv \dot{\bm{X}})$:
\begin{equation}
  \bm{0} = \bm{F}_{\mathrm{H}}(\bm{X},\bm{U})
  +\bm{F}_{\mathrm{R}}(\bm{X})+ \bm{F}_{\mathrm{C}}(\bm{X}).
  \label{eq:force_balance}
\end{equation}
The hydrodynamic forces consist of two components,
a drag due to the motion relative to the surrounding fluid,
$-\bm{R}_{\mathrm{FU}}(\bm{X}) \cdot \bigl(\bm{U}-\bm{U}^{\infty} \bigr)$,
and a resistance to the deformation imposed by the flow,
$\dot\gamma\bm{R}_{\mathrm{FE}}:\hat{\bm{E}}^{\infty}$,
where $\bm{U}^{\infty}_i = \dot\gamma y_i \bm{\hat{e}}_x$
and
$\hat{\bm{E}}^{\infty}
\equiv (\bm{\hat{e}}_x \bm{\hat{e}}_y +\bm{\hat{e}}_y\bm{\hat{e}}_x)/2$
is the normalized strain rate tensor.
(In very dense suspensions, the use of $\bm{U}^{\infty}$  to model the effect of the surrounding fluid is unclear~\citep{Ball_1997}. 
$\bm{U}^{\infty}$ tends to reduce the
fluctuations around the homogeneous flow, but we have established that our results
are unchanged if we remove this term, which is much smaller than the
lubrication and contact forces.)
The resistance matrices $\bm{R}_{\mathrm{FU}}$ and
$\bm{R}_{\mathrm{FE}}$ contain the Stokes drag and the leading terms
of the pairwise hydrodynamic lubrication interaction regularized
to allow contacts~\citep{Jeffrey_1984,Mari_2014}.
Contacts are modeled by a linear spring consisting of both normal and tangential components, with
spring constants $k_{\mathrm{n}}$ and $k_{\mathrm{t}}$, a simple model
commonly used in granular physics~\citep{Mari_2014}.
In contrast to most granular physics models, however, there is no dashpot
in our contact model, as the energy dissipation is provided by
the hydrodynamic resistance.
The electrostatic repulsion decays exponentially with the
interparticle gap $h$ as
$|\bm{F}_{\mathrm{R}}|= F^{\ast} \exp(-h/\lambda_{\mathrm{D}})$, with a Debye
length $\lambda_{\mathrm{D}}$.


The equation of motion~\eqref{eq:force_balance} is completed by the
constraint of flow at constant shear stress $\sigma$.
At any time, the stress in the suspension is given by:
\begin{equation}
  \label{eq:stress}
  \sigma = \dot\gamma \eta_0\biggl(1+\frac{5}{2}\phi\biggr)
  + \dot\gamma \eta_{\mathrm{E}}+\sigma_{\mathrm{R}}+ \sigma_{\mathrm{C}}
\end{equation}
with $\eta_0$ the suspending fluid viscosity,
$\eta_{\mathrm{E}} = V^{-1} \bigl\{ (\bm{R}_{\mathrm{SE}}
-\bm{R}_{\mathrm{SU}}\cdot\bm{R}_{\mathrm{FU}}^{-1}
\cdot\bm{R}_{\mathrm{FE}} ) :\hat{\bm{E}}^{\infty}\bigr\}_{xy} $
and
$\sigma_{\mathrm{R,C}} = V^{-1}\big( \bm{X}\bm{F}_{\mathrm{R,C}}
-\bm{R}_{\mathrm{SU}}\cdot\bm{R}_{\mathrm{FU}}^{-1}\cdot\bm{F}_{\mathrm{R,C}}
\big)_{xy}$, where $\bm{R}_{\mathrm{SU}}$
and $\bm{R}_{\mathrm{SE}}$ are resistance matrices
giving the lubrication stresses from the particles' velocities
and resistance to deformation, respectively~\citep{Jeffrey_1992,Mari_2014}
and $V$ is the volume of the simulation box.
At fixed shear stress $\sigma$ the shear rate $\dot\gamma$ is the fluctuating variable
that is to be determined at each time step by
\begin{equation}
  \label{174314_1Nov14}
  \dot{\gamma} = \frac{\sigma - \sigma_{\mathrm{R}}-
    \sigma_{\mathrm{C}}}{\eta_0\Bigl(1+2.5\phi \Bigr) + \eta_{\mathrm{E}}}.
\end{equation}
The full solution of the equation of motion \eqref{eq:force_balance}
under the constraint of fixed stress \eqref{eq:stress}
is thus the following velocity $\bm{U}$:
\begin{equation}
 \bm{U}  = \bm{U}^{\infty}(\dot{\gamma})
  +
  \bm{R}_{\mathrm{FU}}^{-1}\cdot
  \bigl(
  \dot\gamma
  \bm{R}_{\mathrm{FE}}:\hat{\bm{E}}^{\infty}  
  + \bm{F}_{\mathrm{R}}
  + \bm{F}_{\mathrm{C}}
\bigr).
\label{174328_1Nov14}
\end{equation}
We emphasize here that the stress control is
\emph{deterministic}, in the sense that the shear rate can be
computed \textit{a priori}, before applying the time step.
This is a substantial advantage over the common technique of
stress control using feedback, which adapts the shear rate based on
the results of previous time steps.
In particular, we do not have to introduce the
relaxation time of the feedback loop as an additional time scale, and the shear stress of the
system is strictly constant along the simulation.
Boundary effects like layering are avoided by the use of periodic boundary conditions.
It is also interesting to note that the time
evolution~\eqref{174314_1Nov14} and \eqref{174328_1Nov14} is valid even in a
jammed state: we can have $\dot\gamma=0$ and $\bm{U} = \bm{0}$ if
$\sigma = \sigma_{\mathrm{R}} + \sigma_{\mathrm{C}}$ and
$\bm{F}_{\mathrm{R}} + \bm{F}_{\mathrm{C}} = \bm{0}$.
Lastly, the unit scale is $\dot{\gamma}_0 \equiv F^{\ast} /6 \pi \eta_0 a^2 $
for the strain rate and $\sigma_0 \equiv \eta_0 \dot{\gamma}_0$ for the stress.

\begin{figure*}[!th]
  \centering
  \includegraphics[width=0.92\textwidth]{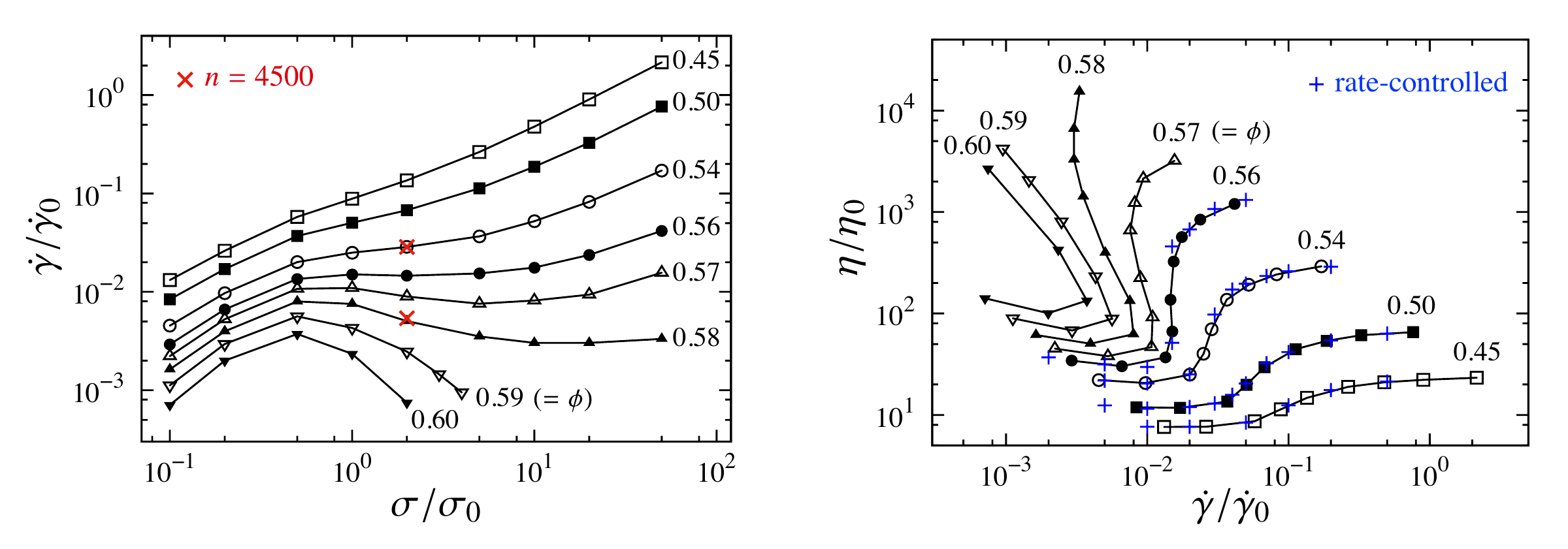}
 \caption{(Color online)  
  \textbf{Left}:
  The shear rate $\dot\gamma$ as a function of the shear stress
  $\sigma$ for several values of the volume
  fraction $\phi$ in our simulations at constant shear stress.
  Continuous shear thickening at the lowest $\phi$ is associated with monotonic flow curves, discontinuous shear thickening appears here as an S-shaped flow curve for $\phi=0.56$--$0.58$, and shear jamming is indicated by arch-shaped flow curves with a vanishing shear rate at high
  stresses for $\phi\geq 0.59$.
  Simulations are run with $n=500$ particles except for $\phi = 0.58$,
  for which we use $n=2000$.
  Two data points with $n=4500$, corresponding to the systems shown in \figref{fig:udiff}, are indicated by red crosses.
  \textbf{Right}:
  The same data plotted as
  viscosity $\eta$ versus shear rate $\dot\gamma$
  are compared with simulations at constant shear rate
  (blue crosses).
}
\label{fig:rate_vs_stress}
\end{figure*}

\section{Results}

In \figref{fig:rate_vs_stress},
we show the average shear rate as a function of
the shear stress for several volume fractions $\phi$.
At the lowest values of $\phi$ the $\dot\gamma(\sigma)$ curves are
monotonic and show a continuous shear thickening for $1 \lesssim \sigma/\sigma_0 \lesssim 10$.
The flow curve becomes non-monotonic at $\phi=0.56$, and for $\phi=0.57$ and $0.58$ the negative slope for intermediate stresses is
clearly visible, showing that an S-shaped rheology curve lies behind the discontinuous shear
thickening observed at the same volume fraction under a rate controlled
simulation. 
Lastly, for $\phi> 0.58$, the flow curves are arches and
the system flows only for small values of the stress, as the shear rate vanishes above a stress
$\sigma_{\mathrm{max}}(\phi)$.
For $\sigma>\sigma_{\mathrm{max}}(\phi)$ the system cannot flow; it is
in a shear jammed state, similar to the one observed in granular
systems~\citep{Bi_2011}.
In our shear jammed states, the entire stress is supported by the repulsive and
contact forces and particles have vanishing
velocities.
Unlike in a dry granular material, however, the system flows
again in steady state with a subsequent decrease in stress for $\sigma<\sigma_{\mathrm{max}}(\phi)$.
(There is also a yield stress above which the
system flows again, due to the deformation of the particles under
shear, but this stress scale is significantly larger than $\sigma_{\mathrm{max}}(\phi)$
and
we have not probed it.)
This phenomenology is also present for the viscosity $\eta$
as a function of $\dot\gamma$
plotted in the right side of \figref{fig:rate_vs_stress},
which shows the same non-monotonicity for $\phi\gtrsim 0.56$.
The calculations at constant shear rate are also shown, and the two
sets agree very well out of the DST transition zone.

\begin{figure}[htbp]
\centering
\includegraphics[width=0.48\textwidth]{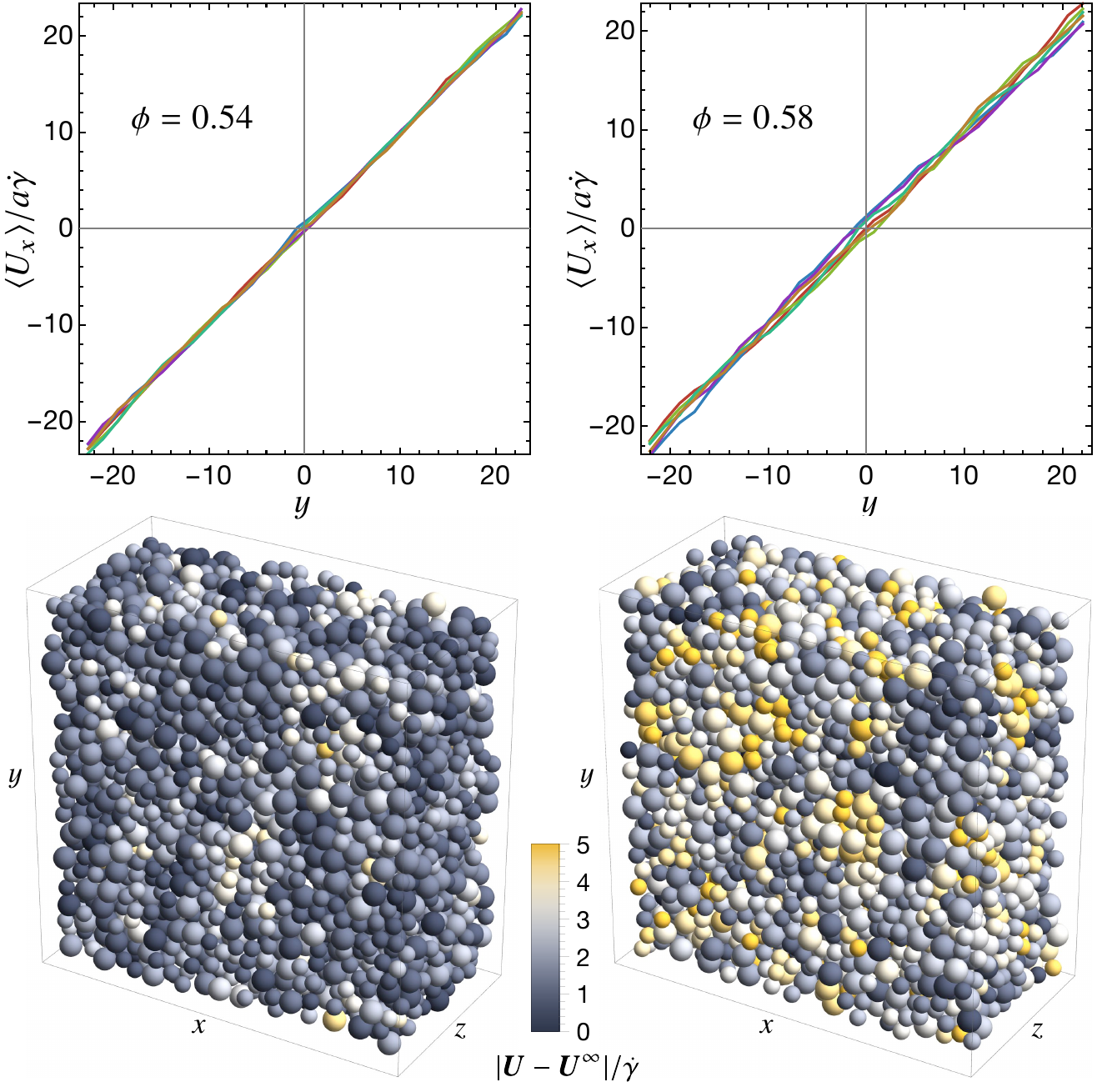}
\caption{(Color online)  \textbf{Top}: Instantaneous velocity
  profiles in the shear thickening regime ($\sigma/\sigma_0=2$) for
  $\phi=0.54$ (left), where $\mathrm{d}\dot\gamma/\mathrm{d}\sigma>0$,
  and for $\phi=0.58$ (right), where
  $\mathrm{d}\dot\gamma/\mathrm{d}\sigma<0$.
  The different curves are taken at different strains separated by one strain unit.
  In both cases the shear flow is uniform.
  \textbf{Bottom}: Configurations of
  the system at $\phi=0.54$ (left) and
  $\phi=0.58$ (right) and stress $\sigma/\sigma_0=2$ as above, color coded with the norm of the non-affine velocity, form
  dark gray (small value) to yellow (light gray, large value). Fluctuations are larger at
  higher volume fraction, but they do not show any sign of non-uniform
  flow or phase separation.}
\label{fig:udiff}
\end{figure}

\begin{figure}[htbp]
\centering
\includegraphics[width=0.48\textwidth]{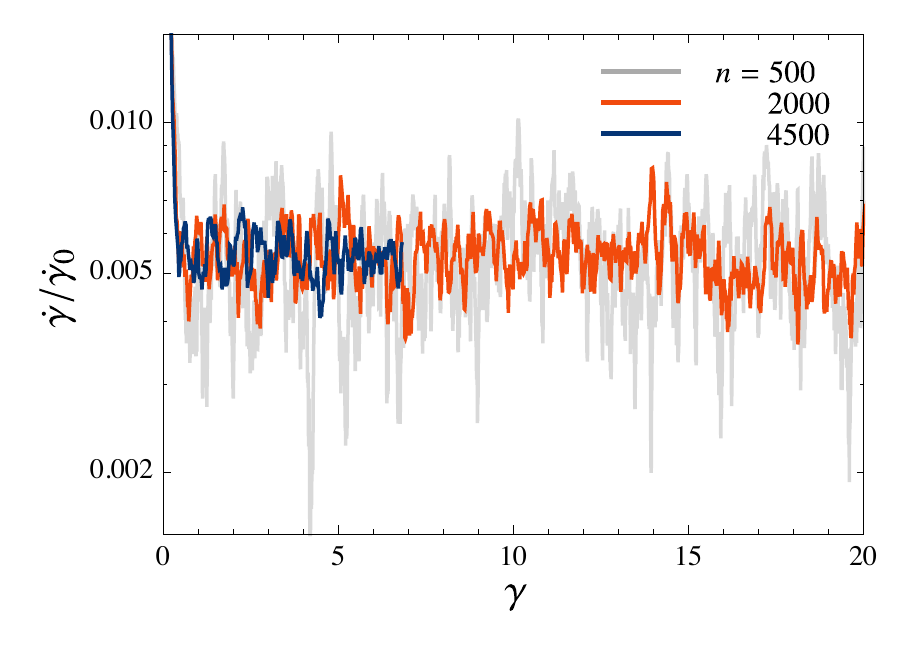}
\caption{(Color online)  Strain series of the shear rate for $\phi=0.58$ and
  $\sigma/\sigma_0=2$, for which the steady state flow curve
  shows $\mathrm{d}\dot\gamma/\mathrm{d}\sigma<0$. The system shows no
  intermittency or exotic fluctuating or chaotic behavior and is in a
  simple steady state. Three sizes are shown: $n=500,2000,4500$, from
  which we can see that the variance of the signal decreases with
  increasing $n$, as is expected for a steady uniform shear
  flow. }\label{fig:shearrate_variance}
\end{figure}

The system shows no sign of spatial heterogeneity within the
$\mathrm{d}\dot\gamma/\mathrm{d}\sigma<0$ region, and the uniform
shear flow is preserved.
In the upper panel of \figref{fig:udiff}
we plot the instantaneous velocity profile at a stress $\sigma/\sigma_0=2$ for $\phi=0.54$ and $\phi=0.58$; both are in the shear
thickening regime, but with
$\mathrm{d}\dot\gamma/\mathrm{d}\sigma>0$ for $\phi=0.54$ and
$\mathrm{d}\dot\gamma/\mathrm{d}\sigma<0$ for $\phi=0.58$.
The velocity profile is linear across the sample in both cases,
indicating a uniform flow.
The normalized fluctuations around the mean flow at particle level,
$|\bm{U}_i-\bm{U}^{\infty}_i|/\dot\gamma a$, are shown as colorcoded snapshots in the lower panel of \figref{fig:udiff}. 
These normalized fluctuations are larger for the denser $\phi=0.58$
than for $\phi=0.54$ (they may actually diverge with the viscosity at the
jamming transition, as a dissipation argument shows~\citep{Ono_2003}),
but they do not show any structure associated with a phase separation.
Neither do they show large fluctuations in the time series, as has
been observed in some experiments~\cite{Frith_1996}.
We show the strain series of the shear rate for three system sizes in
the negative slope region in \figref{fig:shearrate_variance}.
The variance of the shear rate decreases with increasing system size
$n$, and there is no sign of intermittency or macroscopically chaotic
behavior.
This indicates that the uniform shear is the steady state in this region.

\section{Stability analysis}

For micellar systems, where one can observe a similar decreasing flow
curve, the uniform flow is mechanically unstable against
perturbations of the velocity field along the flow
direction due to inertia if $\dot\gamma$ and
$\sigma$ are instantaneously related through a decreasing flow curve~\citep{Spenley_1996,Olmsted_1999,Olmsted_2008,Coussot_2014}.
Our simulations are strictly inertialess, so it is natural to address
the question of the validity of our results for the small but finite
inertia observed in experiments.
We carry out a linear stability analysis at a simplified scalar level
for a suspension at a volume fraction $\phi$ and shear stress $\sigma$
within the decreasing region of the flow curve, i.e., where $\mathrm{d} \dot\gamma / \mathrm{d} \sigma <0$..
We show that the system is linearly stable to perturbations along the
flow direction for small Reynolds numbers and finite system sizes
because of a delay inherent to the microstructural reorganization in
the suspension.
(This analysis is close in spirit to the analysis provided
by~\citet{Nakanishi_2012} for an effective hydrodynamic model of shear
thickening fluids, and it leads to similar conclusions.)

We perform a stability analysis of the momentum equation:
\begin{equation}
  \label{eq:navier_tensor}
  \rho \left(\frac{\partial \bm{v}}{\partial t} +\bm{v} \cdot \nabla \bm{v} \right)= \nabla \cdot \bm{\Sigma}
\end{equation}
that we restrict to simplified scalar (one-dimensional) perturbations.
The unperturbed velocity profile is uniform 
$\bm{v}(\bm{r})  = \dot{\gamma}^{\ast} y\bm{\hat{e}}_x$
and it is associated with a constant shear stress field
$\bm{\Sigma}_{xy}(\bm{r}) = \sigma^{\ast}$.
We introduce a perturbation $\delta v(y,t)\bm{\hat{e}}_x$
in the velocity field and $\delta \sigma(y,t)$ in the stress field.
As $\bm{v} \cdot \nabla \bm{v} = 0$, the momentum equation becomes:
\begin{equation}
  \label{eq:navier}
  \rho \frac{\partial \delta v}{\partial t} = \frac{\partial \delta \sigma}{\delta y}.
\end{equation}

We then need a relation between the $\sigma$ and $\dot{\gamma}$ that is valid at any time.
If these quantities are linked through
their steady state relation $\dot{\gamma}^{\ast}(\sigma)$
(that is, if the relaxation time is strictly zero),
the uniform flow is unstable whenever
$\mathrm{d} \dot{\gamma}^{\ast}/ \mathrm{d} \sigma <0$~\citep{Coussot_2014}.
However, these two quantities are actually linked through the microstructure of the system:
the number of frictional contacts and their anisotropy directly affect
the relation between $\dot{\gamma}$ and $\sigma$.
When there is a local perturbation the microstructure response requires
strain to evolve.
To a first approximation, we can account for this microstructural effect through a
scalar variable $f$, the fraction of frictional contacts,
such that at any time the viscosity (at fixed volume fraction) is a
unique function $\eta(f)$~\citep{Wyart_2014,Mari_2014}.
Building or destroying the contact network takes strain,
and $f(y,t)=f^{\ast}+\delta f(y,t)$ can
be assumed to evolve in the linear regime as
\begin{equation}
  \label{eq:deltaf_eq}
  \delta f(y,t) =
  \int_{-\infty}^{\gamma(t)}
  \mathrm{d}\gamma' K(\gamma-\gamma') \delta \sigma(y,\gamma'),
\end{equation}
where $ K $ is a memory kernel that
depends on the initial steady state stress $\sigma^{\ast}$.
Changing the variable of integration to time and using the fact that
$\delta \sigma = \delta\eta \dot{\gamma}^{\ast} + \eta^{\ast} \delta \dot{\gamma}$,
we then have at linear order:
\begin{equation}
  \label{eq:stress_evo}
  \delta \sigma = \eta^{\ast} \delta \dot{\gamma}
  + \dot{\gamma}^{\ast 2} \frac{\mathrm{d}\eta^{\ast}}{\mathrm{d}f}
   \int_{-\infty}^{t}\mathrm{d}t' K (\dot{\gamma}^{\ast}(t-t'))
   \delta\sigma(t') .
\end{equation}


Now, assuming a periodic perturbation
$\delta v(y, t) = w(t) e^{i k y}$ and $\delta \sigma(y, t) = r(t) e^{i k y}$,
taking the Laplace transform of \eqref{eq:stress_evo} and \eqref{eq:navier},
we have:
\begin{equation}
  \hat{r}(s) =
  \frac{ ik\eta^{\ast} w(0)}{ s+ k^2 \eta^{\ast}/\rho - s\dot{\gamma}^{\ast} (\mathrm{d} \eta^{\ast}/\mathrm{d} f) \hat{K} (s/\dot{\gamma}^{\ast})}
\end{equation}
with $\hat{r}(s)$ and $\hat{K}(s)$ the Laplace transforms of $r(t)$ and $K(t)$.
The final value theorem states that $r(t)$ vanishes when
$t \to \infty$ if $\hat{r}(s)$ has no singularity in the right half of the complex plane,
that is, if
$s+ k^2\eta^{\ast}/\rho - s\dot{\gamma}^{\ast}
(\mathrm{d}\eta^{\ast}/\mathrm{d} f)
\hat{K}(s /\dot{\gamma}^{\ast})$ has no roots
with positive real part.


In the simple case of an exponential kernel
$K(t)= K' e^{-c\dot{\gamma}^{\ast}t}$
(with $K'= c\, \mathrm{d}f^{\ast}/ \mathrm{d}\sigma$), 
we are thus looking at solutions of:
\begin{equation}
  \label{eq:roots_eq}
  s+\frac{k^2\eta^{\ast}}{\rho} - \dot{\gamma}^{\ast} c
  \frac{\mathrm{d}\eta^{\ast}}{\mathrm{d} \sigma}
  \frac{s}{s/\dot{\gamma}^{\ast}+c} = 0,
\end{equation}
which is a quadratic equation
with solutions $s_1$ and $s_2$ satisfying:
\begin{equation}
  \label{eq:sol_prod}
   s_1s_2 > 0,
   \qquad
   s_1+s_2 = -\eta^{\ast}\dot{\gamma}^{\ast}
   c \frac{\mathrm{d}\dot{\gamma}^{\ast}}{\mathrm{d}\sigma}
   - \frac{k^2\eta^{\ast}}{\rho},
\end{equation}
meaning that the real parts of both roots are
negative if
$\mathrm{d}\dot{\gamma}^{\ast} / \mathrm{d}\sigma >-k^2/(\rho
\dot{\gamma}^{\ast} c)$. However, when
$ \mathrm{d}\dot\gamma^{\ast} / \mathrm{d}\sigma <-k^2/(\rho
\dot{\gamma}^{\ast} c)$, the roots have a positive real part,
which means that the uniform shear flow is unstable. 
This stems from the
competition between the time needed to build a microstructure
$(\dot{\gamma}^{\ast} c)^{-1}$
(the longer the more stable) and the
damping time $\rho/(k^2\eta^{\ast})$ on a length scale $k^{-1}$ (the longer
the more unstable).
In the thermodynamic limit the wave-vector $k$ can take arbitrarily
small values and the uniform shear flow will be unstable whenever
$ \mathrm{d}\dot{\gamma}^{\ast} / \mathrm{d}\sigma<0$.
However if the flow occurs in a finite gap $H$
the smallest wave vector available is $k=\pi/H$ and the
uniform shear flow is stable for Reynolds numbers
$\ReNum \equiv \rho\dot{\gamma}^{\ast} H^2/\eta^{\ast}$ smaller than a
critical $\ReNum_{\mathrm{c}}$:
\begin{equation}
 \mathrm{Re_c} = -\frac{\pi^2}{c\eta^{\ast}}
  \left(\frac{\mathrm{d}\dot{\gamma}^{\ast}}{\mathrm{d}\sigma}\right)^{-1}.
\end{equation}
Note that for the simulations shown in this article, the stability is
provided not by the finite size of the system, but by the fact that
the inertia is strictly neglected and $\ReNum=0$.


\newcommand\gammaStepIncrease{\gamma^{\mathrm{step}}}

\begin{figure}
\centering
\includegraphics[width=0.48\textwidth]{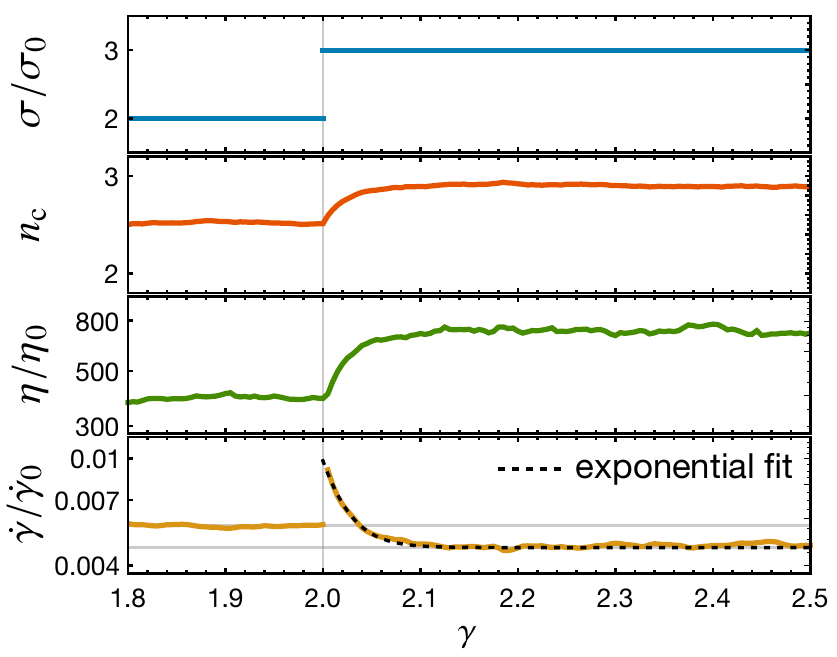}
\caption{(Color online) Typical strain response to a step increment of applied stress
  in the region
  $\mathrm{d}\dot{\gamma}^{\ast}/\mathrm{d}\sigma <0$. These
  curves are obtained by averaging $100$ simulations at $\phi=0.58$, for
  which we impose a stress $\sigma/\sigma_0=2$ for strains
  $\gamma<2$ and increment to $\sigma/\sigma_0=3$ at $\gamma=2$ (top panel).
  The average contact number per particle $n_{\mathrm{c}}$ and
  the viscosity $\eta$ increase (middle panels),
  with an exponential-like relaxation,
  but the shear rate $\dot{\gamma}$ has a non-monotonic behavior,
  first increasing and then relaxing to a lower value (bottom panel).
  The dashed line is a fit to an exponential relaxation,
  giving a relaxation strain of $c^{-1}\approx 0.023$ for these conditions.
  Hence at short time scales, and in particular on the inertial time scale, the
  shear rate increases with the shear stress and the system is
  mechanically stable.
}
\label{fig:step_rheo}
\end{figure}

In order to understand this stability from a physical point of view,
we can consider the case of a step increase of the shear stress
at a given strain $\gammaStepIncrease$,
so that $\sigma(\gamma)=\sigma^{\ast}+\Theta(\gamma-\gammaStepIncrease)\delta\sigma$,
for a stress $\sigma^{\ast}$ in the region
$\mathrm{d}\dot{\gamma}^{\ast} / \mathrm{d}\sigma<0$.
Using \eqref{eq:navier}-\eqref{eq:stress_evo} and , we find that whereas the
local microstructure variable 
and the viscosity
respond monotonically to the shear stress increase, the shear rate
does not: it first increases before relaxing to its steady state value as
$\dot{\gamma}(\gamma) = \dot{\gamma}^{\ast} +
\left[(\mathrm{d}\dot{\gamma}^{\ast}/\mathrm{d}\sigma)
\Bigl(1-e^{-c(\gamma-\gammaStepIncrease )} \Bigr)
+e^{-c(\gamma-\gammaStepIncrease)}/\eta^{\ast}
\right]\Theta(\gamma-\gammaStepIncrease)\delta\sigma$.
As shown in \figref{fig:step_rheo}, this nonmonotonic response is
observed in our simulations: following a step increase in stress, the shear rate, initially at
$\dot{\gamma}^{\ast}$, immediately jumps to a higher value before
decreasing to its steady state value
$\dot{\gamma}<\dot{\gamma}^{\ast}$ in a relaxation that is
reasonably fit by an exponential decay with strain scale
$c^{-1}=0.023$ (i.e. $\SI{2.3}{\percent}$).
This can be understood if we decompose the stress response: the stress
increase in the bulk just after the perturbation at
$\gamma=\gammaStepIncrease$ cannot have its origin from contacts,
because the microstructural reorganization needed to accommodate the
stress change through a contact network takes time to build up.
The initial stress response is of hydrodynamic origin, and this
component of the stress increases only upon an increase of the shear rate.
As a consequence, if the Reynolds number is small enough, on the
inertial time scale the suspension always behaves like a stable fluid
that flows faster with increased applied stress.
%


\section{Discussion}

Our numerical simulations and the linear stability analysis indicate
that it may be possible to observe the S-shaped rheology in experiments,
and S-shaped flow curves have recently been observed for a neutrally-buoyant suspension
of spheres~\citep{Pan_2014}, although the simulation does not capture the
hysteresis and rate-dependent onset of the re-entrant portion of the
flow curve seen in these experiments.
In some stress-controlled experiments the DST shows up in
the $\dot\gamma(\sigma)$ curve as an intermediate plateau with
$\mathrm{d}\dot\gamma/\mathrm{d}\sigma=0$.
The Reynolds numbers involved in these experiments are often within
the range for which our stability analysis predicts a stable uniform
shear flow, so deep stable S-shaped flow curves would be noticed in
the experimental data.  
For polymer beads in glycol~\citep{Laun_1994}
and comb polymer coated PMMA beads in various organic
solvents~\citep{Frith_1996} at volume fractions similar to the ones
studied in this article, experiments show first a decrease and then a
plateau for the shear rate above the discontinuous shear thickening
onset stress.
For precipitated calcium carbonate suspensions, a linear decrease of
the shear rate is sometimes observed~\citep{Egres_2005,Nenno_2014}.
A recent experiment on fluidized gypsum suspensions in
water~\citep{Neuville_2012} shows the arching flow curves associated
with shear jamming resembling the ones we obtain in this work for $\phi>0.58$.
It then seems that the decreasing flow curve is stable under some conditions,
but that most of the time it is hidden by another phenomenon leading to a shear rate plateau.


Our stability analysis is highly restrictive, and the absence of
non-monotonic flow curves in some experiments could be a consequence
of another type of instability. 
To this end, a stability analysis
extended to at least two dimensions would be valuable, but for now we
are limited by the rather simplistic constitutive connection between
the microstructure and the rheology. 
Furthermore, the ``order parameter'' $f$ is a scalar, and we neglect the volume fraction
field and its fluctuations.
Finally, another important point may be the neglect of the small but
finite macroscopic elasticity, which can stem from a conservative
interaction between particles or a finite Brownian motion: this
elasticity can sometimes cause an instability~\citep{Cates_2002}.
%

\section*{Acknowledgments}

MMD is happy to acknowledge a collaboration with Daniel Bonn, whose unpublished experimental data were an important stimulus to this work. 
Our code makes use of the CHOLMOD library by Tim Davis for direct Cholesky factorization of the sparse resistance matrix~\citep{Cholmod}.
This research was supported in part by a grant of computer time from the City University of New York High Performance Computing Center under NSF CNS-0855217, CNS-0958379, and ACI-1126113. 
J.F.M. was supported in part by NSF PREM (DMR 0934206).

\end{document}